\documentclass[conference]{IEEEtran}
\IEEEoverridecommandlockouts

\PassOptionsToPackage{hyphens}{url}

\usepackage{cite}
\usepackage{amsmath,amssymb,amsfonts}
\usepackage{graphicx}
\usepackage{booktabs}
\usepackage{multirow}
\usepackage{array}
\usepackage{xcolor}
\usepackage[hypertexnames=false]{hyperref}
\usepackage{url}
\usepackage{xurl}
\usepackage{algorithm}
\usepackage{algpseudocode}
\usepackage{bm}
\usepackage{mathtools}
\usepackage{orcidlink}
\usepackage{makecell}

\allowbreak

\newcommand{\orcidYunqiWang}{0000-0003-1013-5497}
\newcommand{\orcidNanduniNimalsiri}{0000-0002-9338-0596}
\newcommand{\orcidGhulamMohyuddin}{0000-0002-3905-4220}
\newcommand{\orcidThomasBrinsmead}{0000-0003-1469-6662}

\hypersetup{
    colorlinks=true,
    linkcolor=black,
    citecolor=black,
    urlcolor=blue
}

\begin{document}
\bstctlcite{IEEEexample:BSTcontrol}

\raggedbottom

\title{Planning-Stage Assessment of Battery-Supported Regional Islanding: A South Australian Case Study}

\author{
\IEEEauthorblockN{Yunqi Wang\,\orcidlink{\orcidYunqiWang}, Nanduni Nimalsiri\,\orcidlink{\orcidNanduniNimalsiri}, Ghulam Mohy-ud-din\,\orcidlink{\orcidGhulamMohyuddin}, and Thomas Brinsmead\,\orcidlink{\orcidThomasBrinsmead}}
\IEEEauthorblockA{
Energy Research Unit, Commonwealth Scientific and Industrial Research Organisation (CSIRO), Australia
}
\thanks{This work is supported by the CSIRO Revolutionary Energy Storage
Systems Future Science Platform (RESS FSP) and Australian Research in Power Systems Transition (AR-PST) programs.}
}

\maketitle

\begin{abstract}
Power systems with high shares of renewable generation may face increased operational vulnerability under regional islanding caused by the loss of major inter-area transmission interfaces. This paper presents a planning-stage transmission-level assessment of regional islanding with battery storage, using a South Australian case study based on a synthetic representation of the Australian National Electricity Market (NEM). An interconnected planning baseline is compared with a stress test scenario involving complete regional-islanding over a 24-hour chronological horizon. The framework quantifies peak unserved power, total unserved energy, and the duration of unserved demand, while characterising battery redispatch and the synchronous inertia accessible to the subsystem. 
Results show that islanding increases both peak and cumulative adequacy shortfalls, with additional stress concentrated in the early and late operating hours, while the duration of shortfall remains unchanged in the representative years. Battery dispatch changes materially under islanding, yet residual adequacy shortfalls remain. The inertia screening further shows that the separated subsystem remains below the adopted indicative threshold throughout all 24 hours in every assessed year. The framework therefore provides a compact basis for identifying regional-islanding vulnerabilities that may be underestimated when planning assessments consider only interconnected operation.
\end{abstract}


\section{Introduction}

The Australian National Electricity Market (NEM) is undergoing a rapid transition toward a low-carbon power system, characterised by the retirement of conventional synchronous generation, the large-scale integration of variable renewable energy (VRE), and the increasing deployment of battery energy storage systems (BESSs). Recent Australian Energy Market Operator (AEMO) reporting indicates that the NEM transition is continuing to accelerate: renewable energy delivery in the NEM has tripled over the past decade and approached 45\% of NEM energy in 2025, while renewables supplied 46.5\% of NEM generation in Q1 2026, the highest first-quarter share on record \cite{AEMO2026ISP,AEMOQED2026Q1}. At the same time, generation is becoming more geographically distributed, with growing reliance on renewable energy zones (REZs) and major transmission corridors to transfer energy across regions \cite{AEMO2024ISP,AEMO2025Methodology}. Under these evolving conditions, the resilience of the NEM to severe transmission disruptions has become an increasingly important operational and planning concern.

A particularly challenging condition arises when the loss of critical transmission corridors leads to regional islanding, resulting in an area becoming electrically islanded from the broader interconnected system \cite{nimalsiri2025power}. Under such conditions, the affected subsystem must rely on the available local generation, storage, and any retained synchronous support to maintain supply--demand balance and frequency stability, and to ensure sustained operation. The practical relevance of this issue has been clearly demonstrated in Australia by the 2016 South Australia blackout, which exposed the vulnerability of a renewable-rich yet weakly interconnected region to severe disturbances and islanding events \cite{Yan2018TPWRS}. More broadly, previous studies have shown that high penetrations of wind and PV can materially alter frequency response characteristics in the Australian power system, especially under low-inertia conditions \cite{Yan2015AppliedEnergy,Gu2018TPWRS}. These challenges are closely linked to evolving security requirements in future power systems. As inverter-based generation replaces conventional synchronous plants, systems increasingly face declining inertia, lower fault-current contribution, and tighter operational margins \cite{Gu2018TPWRS,Gu2020TSTE}. 

In this context, BESS technologies are recognised not only as energy-shifting assets but also as providers of fast-acting support that can enhance system flexibility and frequency-related performance \cite{Akram2020RSER,Zhao2023RSER,AEMOSyntheticInertia2024}. Recent reports published by the AEMO further emphasise inertia adequacy, sub-network requirements, and the role of grid-forming BESS capability in high inverter-based resource (IBR) operation \cite{AEMOInertia2024,AEMOSyntheticInertia2024}. These requirements are particularly relevant under regional islanding, because the separated subsystem can no longer rely on external interconnection support to maintain supply--demand balance, frequency-security margins, and fast post-disturbance response. In this setting, battery-supported islanding is therefore not only an adequacy problem, but also a local security-assessment problem involving retained synchronous support, internal flexibility, and the operational capability of storage resources. Together, these developments make battery-supported regional islanding an increasingly important consideration in future NEM planning and security assessment.

AEMO has increasingly adopted subregional, rather than purely regional, system representations in its planning and market modelling to better capture major intra-regional bottlenecks, demand diversity, and interactions among network development, renewable generation, and storage resources. Although such representations are primarily introduced for long-term planning and development pathway assessment, they also provide a natural basis for studying post-islanding operability in vulnerable parts of the system. Existing transmission-level planning and reduced-network models are typically designed to assess dispatch feasibility, congestion patterns, and augmentation needs under interconnected operating conditions \cite{Wu2006Energy}. Recent work has also demonstrated that synthetic Australian NEM transmission models can support large-scale operational and planning-oriented studies, including security-constrained optimal power flow under evolving AC/DC network conditions \cite{Ghulam2024EPSR}. Their direct application to regional islanding remains comparatively limited, particularly when the objective is to evaluate whether an electrically separated region can sustain operation over time using internal generation and storage resources. Moreover, active-power adequacy alone may be insufficient in such settings, because an islanded subsystem can remain operationally fragile if retained synchronous support is inadequate. Recent work on islanding event-constrained optimal power flow has moved in this direction by explicitly considering selected interconnector loss events and associated inertia-related requirements in reduced Australian system settings \cite{PuschelLovengreen2020EPSR}. Nevertheless, there remains a need for a practical transmission-level framework that applies complete regional separation as an additional stress test to an interconnected planning trajectory and jointly characterises the resulting adequacy shortfalls, battery redispatch, and accessible synchronous inertia.

This paper presents a planning-stage framework for assessing battery-supported regional islanding and demonstrates its application through a South Australian case study in a synthetic NEM setting. 
The baseline planning trajectory is developed for interconnected operation and does not explicitly require South Australia to remain self-sufficient following complete regional separation.
The study therefore applies an islanding stress test to the same annual generation, transmission, and battery portfolio, and evaluates the resulting adequacy shortfalls, battery response, and inertia-screening outcomes over a chronological horizon. Rather than attempting detailed dynamic replication, the paper provides a compact planning-stage basis for interpreting post-islanding operability in a renewable-rich transmission system.

The contributions of this paper are summarised as follows:
\begin{itemize}
    \item We develop a transmission-level framework for stress-testing interconnected power-system planning portfolios under complete regional islanding.
    %
    \item We quantify peak and cumulative adequacy shortfalls and their temporal duration over a multi-period horizon, while characterising how the planned battery fleet is redispatched following regional separation.
    \item We extend the active-power adequacy assessment with supplementary synchronous-inertia screening that compares the inertia accessible during interconnected operation with the local synchronous inertia retained following complete regional separation.
\end{itemize}

\section{Regional Islanding Assessment Framework}

\subsection{Representation of an Islanded Grid}

This paper considers a transmission-level representation of an interconnected power system as a structured platform for the planning-stage assessment of regional islanding. The complete interconnected network represents the baseline operating condition. Regional islanding is modelled by removing the transmission interfaces that connect the selected region to the rest of the system.


Let the transmission network be represented by a graph $G=(V,E)$, where $V$ denotes the set of buses, and $E$ denotes the set of transmission corridors. For a particular islanding event, let $S \subseteq V$ denote the set of buses belonging to the separated subsystem. Complete regional separation is represented by removing all branches in the boundary cut between $S$ and $V$\textbackslash$S$, while retaining the internal network, generation, demand, and battery resources within $S$. The corresponding islanded subsystem is represented by the induced subgraph $G_S=(S,E_S)$, where
$E_S=\{(i,j)\in E \ | \  i\in S,\; j\in S\}$.


Post-islanding operation is assessed on $G_S$ using generation, battery resources, and retained synchronous units within the islanded subsystem, subject to the transfer limits of the remaining internal transmission network. In this paper, the problem is interpreted as a chronological adequacy screening of the isolated subsystem rather than a detailed operational or dynamic replay of the full interconnected system. This reduced representation preserves the internal topological structure of the islanded region while remaining tractable for multi-year screening.


\subsection{Post-islanding Adequacy Metrics}

In this paper, system adequacy refers to the ability of the islanded subsystem to meet local demand over the assessment horizon using available generation and battery resources within the islanded subsystem, subject to the transfer limits of the remaining internal transmission network. Under the baseline scenario, these transfers include power exchanged through the external interconnectors. Under the islanding scenario, the subsystem must rely solely on its internal resources and transmission network.

Assessing adequacy is particularly important under islanding because the separated subsystem can no longer rely on imports or balancing support from the wider interconnected network. Any residual power shortfall therefore represents unserved power at a given time step and, when accumulated over time, unserved energy. Such shortfalls may require demand curtailment, emergency operating actions, or additional local flexibility.

We consider a chronological assessment horizon indexed by the time-step set $\mathcal{T}=\{1,\ldots,T\}$, where each time step has duration $\Delta t$. In the case study, $\Delta t$ is one hour and $\mathcal{T}$ represents the hourly intervals of the assessed horizon. For each time step $t\in\mathcal{T}$, the model yields an aggregate positive power-balance slack $p_t\geq0$. This quantity is interpreted as a planning-stage proxy for \textit{unserved power}, representing local supply insufficiency after available generation, battery dispatch, and internal network transfer limits have been accounted for. For concision, this proxy is referred to below as unserved power.


Accordingly, the \textit{peak unserved power} is defined as
\begin{equation}
    P^{\mathrm{peak}} = \max_{t \in \mathcal{T}}p_t.
\end{equation}

The corresponding unserved energy over the study horizon is defined as
\begin{equation}
    E^{\mathrm{uns}} = \sum_{t \in \mathcal{T}} p_t \Delta t.
\end{equation}
Here, $P^{\mathrm{peak}}$ characterises the most severe instantaneous stress under islanding, while $E^{\mathrm{uns}}$ captures the cumulative adequacy impact across the full time horizon.

To characterise the temporal extent of inadequacy, the duration of unserved demand is defined as:
\begin{equation}
    H^{\mathrm{uns}} = \sum_{t \in \mathcal{T}} \mathbf{1}(p_t > 0)\Delta t,
\end{equation}
where $\mathbf{1}(\cdot)$ is the indicator function, which equals $1$ when $p_t>0$ and $0$ otherwise. This metric distinguishes between cases in which post-islanding stress is concentrated in a few critical intervals and those in which supply inadequacy persists over a larger portion of the day.


Taken together, $P^{\mathrm{peak}}$, $E^{\mathrm{uns}}$, and $H^{\mathrm{uns}}$ characterise the peak severity, cumulative impact, and temporal persistence of post-islanding inadequacy.

\subsection{Battery Response Metrics}



Battery operation is represented chronologically to characterise how the planned storage fleet is redispatched under islanded operation. Let $p^{\mathrm{dis}}_{b,t}$ denote the discharge power of battery resource $b$ at hour $t$. The total battery discharge energy over the study horizon is defined as
\begin{equation}
    E^{\mathrm{dis}}
    =
    \sum_{t\in\mathcal{T}}
    \sum_{b\in\mathcal{B}}
    p^{\mathrm{dis}}_{b,t}\Delta t,
\end{equation}
where $\mathcal{B}$ is the set of battery resources within the South Australian subsystem.

To capture the maximum instantaneous battery response, the peak battery discharge power is defined as
\begin{equation}
    P^{\mathrm{dis,peak}}
    =
    \max_{t\in\mathcal{T}}
    \sum_{b\in\mathcal{B}}
    p^{\mathrm{dis}}_{b,t}.
\end{equation}

Let $e_{b,t}$ denote the stored energy of battery $b$ at time $t$. The aggregate battery energy-state swing over the assessed horizon is defined as
\begin{equation}
    \Delta E^{\mathrm{bat}}
    =
    \max_{t\in\mathcal{T}}
    \sum_{b\in\mathcal{B}} e_{b,t}
    -
    \min_{t\in\mathcal{T}}
    \sum_{b\in\mathcal{B}} e_{b,t}.
\end{equation}

The quantities $E^{\mathrm{dis}}$, $P^{\mathrm{dis,peak}}$, and $\Delta E^{\mathrm{bat}}$ characterise cumulative battery discharge, the strongest instantaneous response, and the temporal variation in the aggregate energy state of the planned battery fleet, respectively. These are descriptive battery-response indicators and do not isolate the causal mitigation contribution of storage.

\subsection{Supplementary Inertia Screening}

Because active-power adequacy alone does not guarantee broader operational robustness after islanding, a simplified inertia-based screening metric is included as a supplementary indicator. For each scenario $s\in\{\mathrm{base},\mathrm{isl}\}$, let $\mathcal{C}_s$ denote the AC-connected component containing South Australia. The synchronous inertia accessible to the South Australian subsystem at time $t$ is approximated by
\begin{equation}
    I_{t,s}
    =
    \sum_{g\in\mathcal{G}^{\mathrm{syn}}(\mathcal{C}_s)}
    H_g S_g u_{g,t},
\end{equation}
where $H_g$ is the inertia constant of unit $g$, $S_g$ is its rated apparent power, and $u_{g,t}$ is a dispatch-status indicator that equals one when the synchronous unit is dispatched above a small tolerance.

Under the baseline condition, $\mathcal{C}_{\mathrm{base}}$ includes the wider AC-synchronous component connected to South Australia. Following complete regional separation, $\mathcal{C}_{\mathrm{isl}}$ is restricted to the South Australian island. HVDC links are not treated as providing synchronous inertia coupling.

For this planning-stage screening, a fixed indicative threshold of $\bar I=10{,}000$~MW\,s is adopted for all hours, planning years, and scenarios. This value is used as an illustrative screening reference rather than a formally determined secure operating level. The inertia adequacy ratio is defined as
\begin{equation}
    \Gamma_{t,s}
    =
    \frac{I_{t,s}}{\bar I}.
\end{equation}

The duration of inertia inadequacy is defined as
\begin{equation}
    H^{\mathrm{inad}}_s
    =
    \sum_{t\in\mathcal{T}}
    \mathbf{1}(\Gamma_{t,s}<1)\Delta t,
\end{equation}
and the minimum adequacy ratio is
\begin{equation}
    \Gamma^{\min}_s
    =
    \min_{t\in\mathcal{T}}\Gamma_{t,s}.
\end{equation}

These indicators are not intended to replace detailed dynamic frequency-security analysis. Rather, they provide a compact planning-stage comparison between the inertia accessible under interconnected operation and the retained local inertia following regional separation.

\section{Case Study and Discussion}

\begin{figure}[!bp]
\centering
\includegraphics[width=0.85\columnwidth]{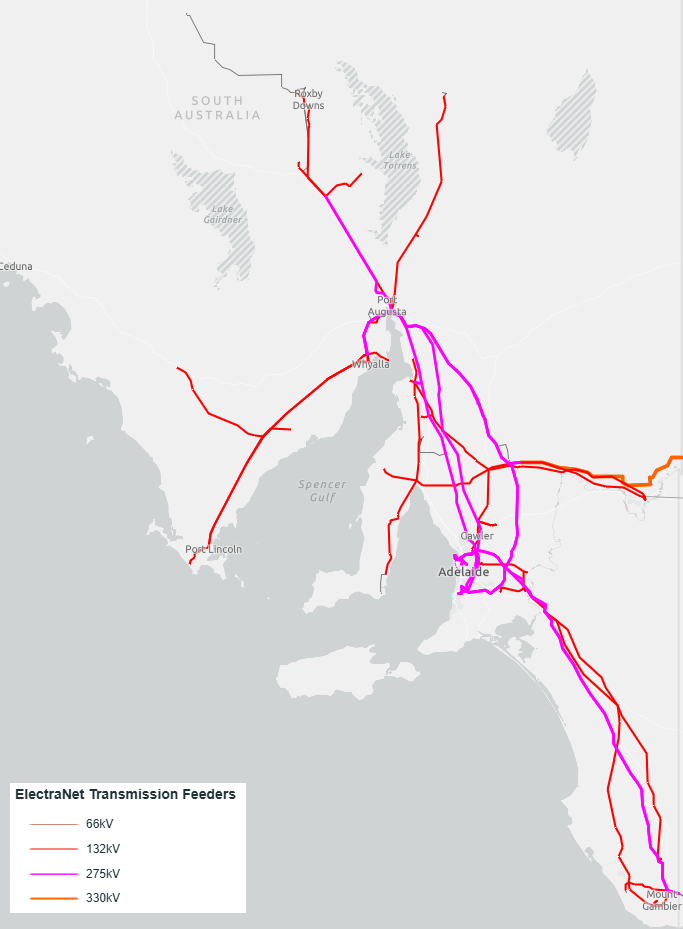}
\caption{South Australian network, showing the major high-voltage corridors in the region. The case study in this paper is based on a reduced transmission-level approximation of the South Australia subsystem within the synthetic NEM setting \cite{electranet_network_map}.}
\label{fig:topology}
\end{figure}

The proposed framework is demonstrated through a South Australian case study within a synthetic NEM setting. The South Australian transmission network, which is owned and operated by ElectraNet, spans a large geographic area and comprises a meshed high-voltage backbone operating at 330, 275, 132, and 66~kV, with more than 90 high-voltage substations and approximately 6600 kilometres of transmission lines. This network supports electricity transfer across a wide and geographically diverse region while linking South Australia to the rest of the NEM through major 
interconnectors, as illustrated in Fig.~\ref{fig:topology}. In this study, South Australia is not modelled as a full engineering replica of the ElectraNet network. Instead, it is represented through a reduced transmission-level approximation within the synthetic NEM. 

The South Australian subsystem is identified from the broader NEM topology and then assessed under an islanding condition created by removing its external AC/DC 
interconnectors linking it to the rest of the system. In this way, the study retains the principal topological role of South Australia as a weakly interconnected regional subsystem, while using a compact representation suitable for repeated planning-stage screening across multiple study years.


\begin{figure}[!tp]
\centering
\includegraphics[width=1.0\columnwidth]{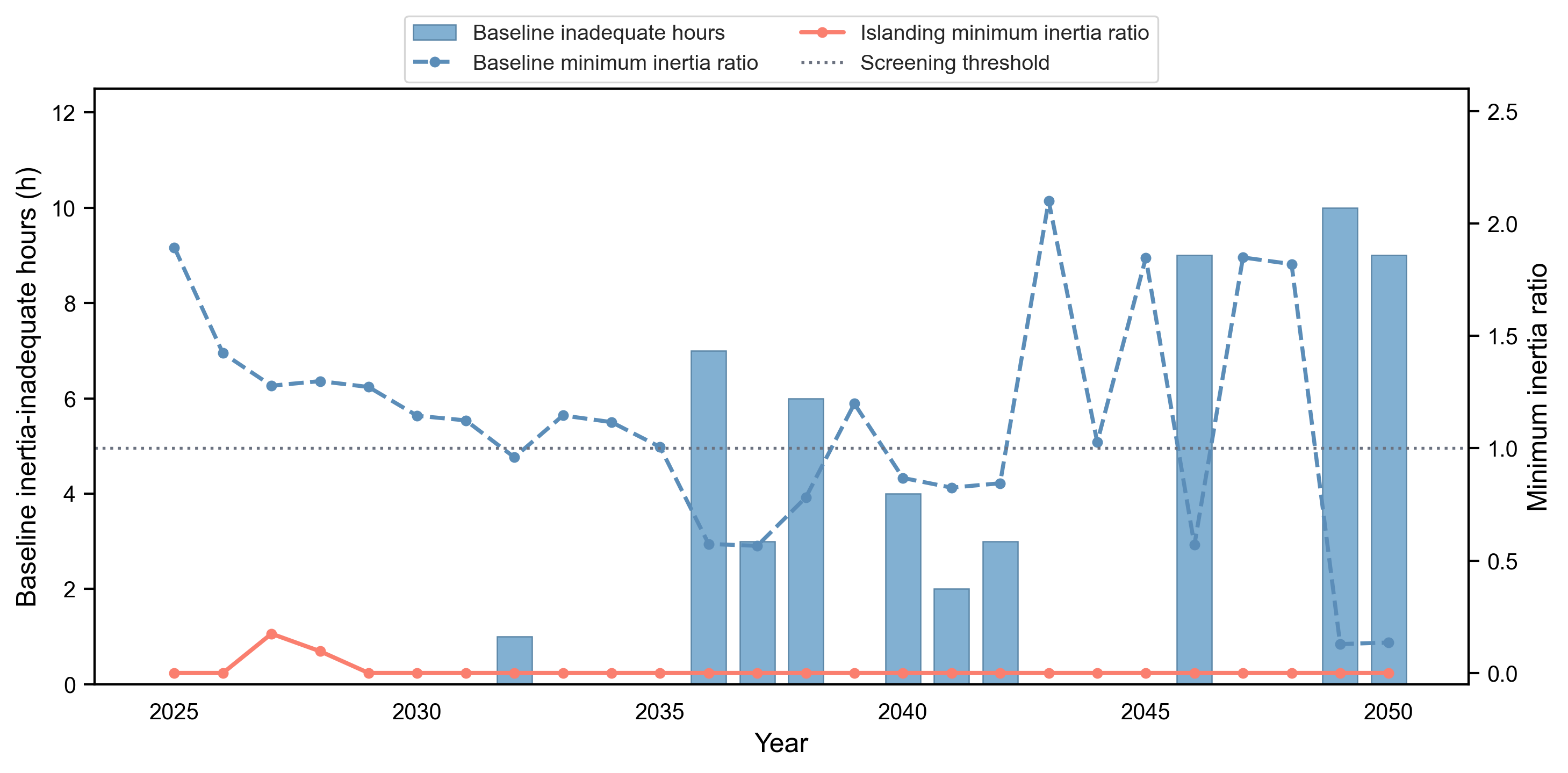}
\vspace{-1em}
\caption{Baseline inertia-inadequate hours and minimum inertia ratios under the baseline and islanding conditions. The islanding case is inertia-inadequate for all daily hours in every assessed year and is omitted from the bar series.}
\label{fig:sa_inertia}
\end{figure}

\begin{figure}[!tp]
\centering
\includegraphics[width=1.0\columnwidth]{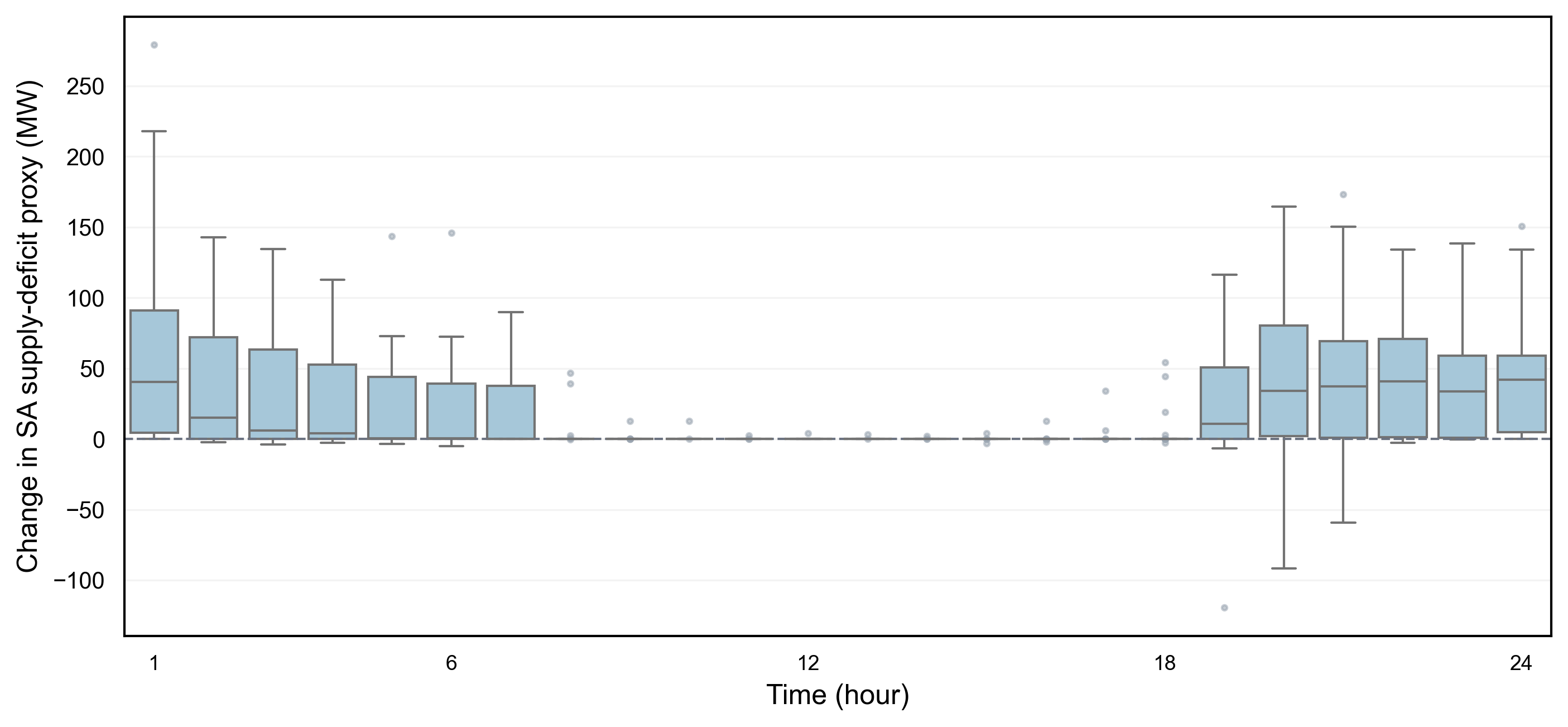}
\vspace{-1em}
\caption{Hourly distribution across years of the islanding-induced change in the South Australian supply-deficit proxy, calculated as the islanding value minus the baseline value.}
\label{fig:sa_unserved}
\end{figure}

\begin{table}[t]
\caption{Representative-year metrics and maximum signed islanding-induced increases.}
\label{tab:sa_metrics_summary}
\centering
\scriptsize
\renewcommand{\arraystretch}{1.12}
\setlength{\tabcolsep}{2.5pt}

\begin{tabular}{lccc}
\hline
\multicolumn{4}{c}{\textbf{Adequacy metrics}} \\
\hline
Scenario
& \makecell[c]{Peak unserved\\power (MW)}
& \makecell[c]{Unserved\\energy (MWh)}
& \makecell[c]{Inertia-inadequate\\hours (h)} \\
\hline
Baseline (2035)   & 605.5  & 5,476.1  & 0  \\
Islanding (2035)  & 637.0  & 5,702.3  & 24 \\
Baseline (2046)   & 1,366.4 & 13,602.6 & 9  \\
Islanding (2046)  & 1,530.9 & 14,972.8 & 24 \\
Max. increase (year)
& 279.1 (2047)
& 1,399.2 (2045)
& 24 (2025) \\
\hline
\multicolumn{4}{c}{\textbf{Battery-response metrics}} \\
\hline
Scenario
& \makecell[c]{Battery discharge\\energy (MWh)}
& \makecell[c]{Peak battery\\discharge power (MW)}
& \makecell[c]{Battery energy-state\\swing (MWh)} \\
\hline
Baseline (2035)   & 21,082.8 & 1,735.0 & 5,269.6  \\
Islanding (2035)  & 20,071.4 & 1,794.8 & 10,493.2 \\
Baseline (2046)   & 30,630.1 & 2,616.0 & 20,357.2 \\
Islanding (2046)  & 20,391.3 & 2,725.1 & 12,585.0 \\
Max. increase (year)
& 8,191.8 (2038)
& 516.7 (2048)
& 6,498.5 (2047) \\
\hline
\end{tabular}
\end{table}

For each planning year from 2025 to 2050, the model is solved over daily chronological horizon at hourly resolution using the same annual generation, transmission, demand, and battery assumptions in both scenarios. The baseline represents interconnected operation and does not explicitly impose regional self-sufficiency following complete separation. The islanding scenario applies an additional planning-stage stress test by removing South Australia's external AC and DC interconnectors, while retaining the same planned resource portfolio. The comparison therefore evaluates whether a portfolio developed for interconnected operation remains adequate following complete regional separation.

Before examining the adequacy impact in detail, it is useful to compare the synchronous inertia accessible to South Australia under interconnected operation with the retained local inertia following complete regional separation. Fig.~\ref{fig:sa_inertia} shows that, under the baseline condition, inertia inadequacy occurs only in selected planning years and becomes more pronounced as the synchronous generation mix evolves. Following islanding, however, the South Australian subsystem remains below the adopted screening threshold for all 24 hours in every assessed year. This persistent outcome shows that the challenge posed by islanding is not purely an active-power adequacy issue, because the planned local synchronous resource mix does not satisfy the adopted indicative threshold once access to the wider AC-synchronous system is lost.

With that broader screening context established, Fig.~\ref{fig:sa_unserved} shows that the islanding-induced change in the supply-deficit proxy is strongly time-dependent. Positive changes are concentrated mainly in the early and late operating hours, whereas the middle of the day remains comparatively less affected. A limited number of negative changes occur for particular year--hour combinations, reflecting differences in system redispatch between the two scenarios. The overall pattern demonstrates that a single operating snapshot may materially understate regional-islanding vulnerability and supports the use of a chronological assessment.

Table~\ref{tab:sa_metrics_summary} quantifies the adequacy impact of complete regional separation in the two representative years. Relative to the interconnected baseline, peak unserved power increases by 5.2\% in 2035 and 12.0\% in 2046, while unserved energy increases by 4.1\% and 10.1\%, respectively. By contrast, the duration of positive supply-deficit conditions remains unchanged at 14 hours in 2035 and 15 hours in 2046. Islanding therefore intensifies shortfalls during operating periods that are already constrained, rather than extending the number of affected hours. Across the full 2025--2050 horizon, the largest signed increases reach 279.1~MW in peak unserved power in 2047 and 1,399.2~MWh in unserved energy in 2045. These results indicate that the principal planning gap lies in the magnitude and concentration of regional adequacy stress, which may not be visible when the same portfolio is assessed only under interconnected operation.

The battery-response metrics reveal a more heterogeneous pattern. In both representative years, peak battery discharge power increases under islanding, by 3.4\% in 2035 and 4.2\% in 2046, indicating a stronger instantaneous response during critical intervals. However, total battery discharge energy decreases by 4.8\% and 33.4\%, respectively. The aggregate battery energy-state swing nearly doubles in 2035, increasing by 99.1\%, but decreases by 38.2\% in 2046. The wider planning horizon also exhibits substantial year-to-year variation: the maximum signed increases in discharge energy, peak discharge power, and energy-state swing are 8,191.8~MWh in 2038, 516.7~MW in 2048, and 6,498.5~MWh in 2047, respectively. These contrasting outcomes show that regional separation reshapes both the timing and depth of storage utilisation rather than uniformly increasing battery use. The battery indicators should therefore be interpreted as evidence of portfolio redispatch, rather than as a direct measure of battery-induced mitigation. Residual adequacy shortfalls remain, showing that the existing interconnected planning portfolio does not ensure self-sustained regional operation.

\section{Conclusion}

This paper presented a planning-stage transmission-level framework for assessing battery-supported regional islanding and demonstrated it through a South Australian case study in a synthetic NEM setting. By comparing an interconnected planning baseline with a complete regional-islanding stress test, the study examined whether the same planned generation, transmission, and battery portfolio remains adequate following regional separation. The results show that islanding increases the magnitude of the South Australian supply-deficit proxy during concentrated early- and late-day operating windows, while the duration of stressed operation remains unchanged in the representative years. Planned battery resources are materially redispatched under islanding, but the existing planning portfolio does not eliminate residual adequacy shortfalls. The supplementary inertia screening further shows that, once access to the wider AC-synchronous system is lost, the South Australian island remains below the adopted indicative threshold for all daily hours in every assessed year.

Overall, the proposed framework provides a compact basis for identifying regional-islanding vulnerabilities that may remain hidden when future planning trajectories are assessed only under interconnected operation.

\bibliographystyle{IEEEtran}
\bibliography{references}

\end{document}